\documentclass[12pt]{article}

\usepackage[lmargin=1.5in, rmargin=1in, tmargin=1in, bmargin=1in]{geometry}
\usepackage{amsmath}
\usepackage{amsfonts}
\usepackage{amssymb}
\usepackage{amsthm}
\usepackage{cancel}
\usepackage{bbm}
\usepackage{setspace}
\usepackage{graphicx}
\graphicspath{ {figures/} }
\usepackage{array}
\usepackage{subcaption}
\usepackage[utf8]{inputenc}
\usepackage[english]{babel}
\usepackage{caption}
\usepackage{float}
\usepackage{afterpage}
\usepackage{geometry}
\usepackage{afterpage}
\usepackage[utf8]{inputenc}
\usepackage[normalem]{ulem} 
\usepackage{listings}
\usepackage[table]{xcolor}
\usepackage{alltt}
\definecolor{maroon}{cmyk}{0,0.87,0.68,0.32}
\theoremstyle{definition}
\newtheorem*{remark}{Remark}
\theoremstyle{case}

\usepackage{appendix}
\usepackage{lscape}
\usepackage[utf8]{inputenc}
\usepackage[affil-it]{authblk}
\usepackage{etoolbox}
\usepackage{lmodern}
\usepackage{multirow}
\usepackage{lineno}

\title{A cellular automata model for particle transport in disordered systems}

\author[1]{Lander Besabe}
\author[2]{Editha Jose}
\author[3]{Alvin Karlo Tapia}

\affil[1]{Department of Mathematics, University of Houston}
\affil[2]{Institute of Mathematical Sciences, University of the Philippines Los Ba\~{n}os}
\affil[3]{Institute of Physics, University of the Philippines Los Ba\~{n}os}

\begin{document}

\maketitle

\begin{center}
    \textbf{Abstract}
\end{center}

\hspace{\parindent}

We construct a cellular automaton (CA) model that describes the movement of a particle in a disordered system. The mathematical properties of the CA model were examined by varying the configuration of grid and determining the number of percolating paths. Through this model, we were able to develop a computer simulation that shows particle transport. Under particle hopping mechanism, with or without tunneling(or backscattering), it was found out that there is an exponential behavior of percolation probability. However, the onset of the percolation probability is shifted to a smaller value when tunneling and backscattering are present. 


\section{Introduction}
\hspace{\parindent}

Studying movement of particles in a medium through experiment is tedious, expensive, and will consume a lot of resources. However, computer simulations will give us meaningful results which will provide both qualitative and quantitative descriptions of the behavior of a particle traveling through a certain medium. The concepts of percolation theory and particle transport will help in this description.

Percolation theory gives us an idea of how particles travel through a medium with barriers and the open-site vacancy requirement or the percolation threshold that will allow the particle to pass through the material \cite{Steif2009ASO}. On the other hand, cellular automata gives a relation on how a certain cell is related to its neighboring cells and how they affect each other \cite{CA}.

The theoretical foundation of the study of cellular automata (CA) was first established around 1952 when John von Neumann modeled self-replicating systems like robots building other robots, and created the first two-dimensional cellular automata, with 29 elementary states. He also proved that the automata will produce a periodic pattern across time. However, earlier work utilized simple lattices by Stanislaw Ulam. Ulam modeled the growth of crystals using simple lattices \cite{history}. 

The cellular automata model \textit{Game of Life}, constructed by John Conway was published in Scientific American by Martin Gardner. The model only consists of two elementary states, dead or alive, but generated a lot of patterns, whether periodic, repeating or turbulent. What strengthened the theories in cellular automata is the work of Stephen Wolfram in 2002. He published \textit{A New Kind Of Science}, which popularized cellular automata to spread across all disciplines of science \cite{history}. 

Since its popularity, researchers from different fields have been enticed to the simplicity of the cellular automata. The concept that the state of a specific cell is dependent on the behavior of its neighbors is applicable to diverse disciplines like biology \cite{BioCA}, fluid mechanics \cite{LeakCA}, pedestrian dynamics \cite{PedestrianCA}, wildfire \cite{wildfireCA}, and many more.

A cellular automata model, based on percolation theory of proton hopping down a channel has been created and studied by Kier et.al in 2013 \cite{LBKier}. Variations in the rate of proton entry into the channel and the effects of the polar character of the channel walls was studied using the model. The behavior of the models corresponds to experimental results. This study used a specific medium and particle. Moreover, it only utilized the concept of cellular automata for the simulation, but did not provide a mathematical model.

Li, Yan, Yang, and Wang in 2016 \cite{ordered} studied how novel particles move in an ordered packed bed. They varied the shape of the packed bed; spherical, and more elliptical three-dimensional models were studied to see its effect on the plate height, which is the square of the standard deviation equal to a constant times the distance traveled.

Cellular automata also plays a big part in the development of machine learning. Kleyko, Frady, and Sommer showed in 2020 that CA can be used to reduce memory requirements when performing collective-state computing, particularly, using the CA90 elementary rule by Stephen Wolfram, et al. They tested it against traditional methods like neural networks and Ising method to verify that using CA90 uses way less memory \cite{Kleyko2020CellularAC}. In 2018, Zhai et al also used agent-based cellular automata to model a machine learning algorithm for the dynamic traffic flow of electric vehicles \cite{evtraffic}.

Most of these studies consider ordered systems. However, we are going to simulate disordered systems. Cellular disorder is the kind of disorder which can be described with reference to a particle placed on an ideal lattice site of a solid. The properties involved are intrinsic, as in the case of spin direction and chemical composition, or pertain to the presence or absence of defects at low or moderate concentration or, as in the cases of thermal motion \cite{disorder}, electrical conductivity \cite{Conduction,compare,disorder}, quantum mechanical systems \cite{quantummech,Nigmatullin2021DirectedPI}, traffic flow \cite{traffic}, and internet science \cite{internet}.

In this study, we construct a CA model for particle transport; develop computer simulation for the CA model and verify using ordered systems; present the mathematical properties of the CA model; and analyze particle transport under different conduction phenomena (with tunneling, backscattering) of disordered systems.

\section{Theoretical Background of the Model}
\begin{figure}[H]
    \centering
    \hspace{-1cm}
    \includegraphics[scale=1.2]{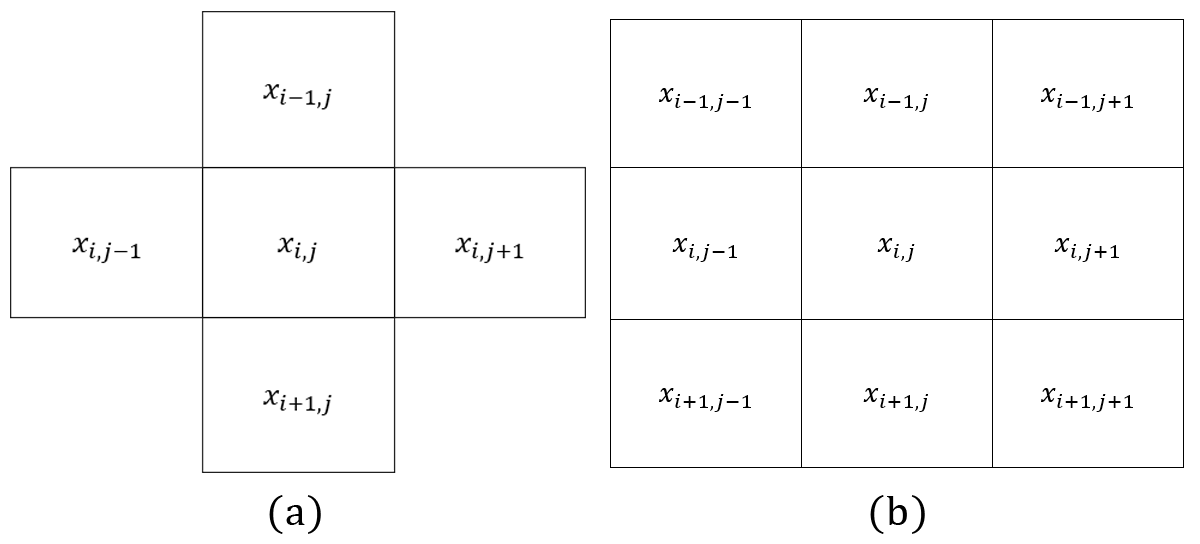}
    \caption{(a) von Neumann Neighborhood and  (b) Moore Neighborhood}
    \label{Neighborhood}
\end{figure}

\hspace{\parindent}
A \textbf{cellular automaton} (CA) is a quadruple $A=(G,E,U,f)$ of a grid  $G$ of cells, a set of elementary states $E$, a set defining the neighborhood $U$, and a local rule $f$. In classical CA, we have $G=\mathbb{Z}^{n}$, as an $n$-dimensional square grid. On the other hand, the elements of the state set $E$ are called states. The state of a cell is dependent on the states of its neighbors. There are two common types of a neighborhood: the von Neumann and the Moore neighborhood. Figure \ref{Neighborhood} shows the difference between the von Neumann and the Moore neighborhood of $x_{i,j}$. 

The \textbf{local rule} (or local update rule, the update rule, or simply the rule) of a CA with state set $S$ and size $m$ neighborhood is a function $$ f:S^{m} \rightarrow S $$
that specifies the new state of each cell based on the states of its neighbors. If the neighbors of a cell have states $s_{1}, s_{2}, \dots, s_{m}$ then the new state of the cell is $f(s_{1}, s_{2}, \dots, s_{m})$. In addition, a configuration assigns a state to each cell in the $d$-dimensional grid $\mathbb{Z}^d$. A configuration serves as a snapshot of each cell at time $t$.

Percolation theory explains how particles flow across the surface of a porous material \cite{IntroToPerc}. This theory also describes how probable particles pass through a material given a percentage of open site vacancy. It has brought new understanding in the field of mathematics and solid-state physics. Several studies on conduction\cite{Conduction}, soil nutrients\cite{soil} and particle transport used percolation theory to describe the behavior of charges \cite{Wang}, fluids and other types of particles.

\begin{figure}[H]
    \centering
    \includegraphics[scale=0.75]{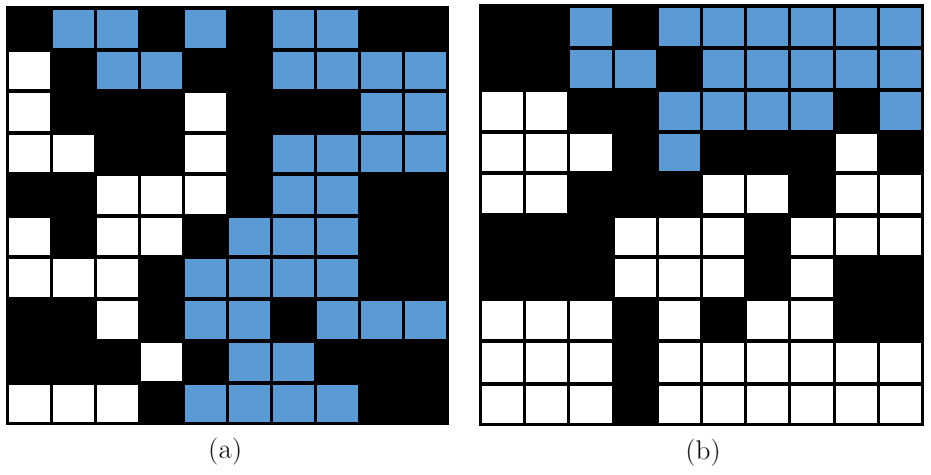}
    \caption{Illustration of (a) percolating and (b) non-percolating system}
    \label{Percolation}
\end{figure}

Figure \ref{Percolation} shows the static percolation in a material of a certain open site vacancy. In the lattice below, white cells represent open sites, meaning the particles can move to those cells; black cells represent blocked sites or the cells where the particles cannot pass through; and blue cells represent filled sites, or sites where the particles are. Figure \ref{Percolation} (a) shows that the particles successfully percolated, while Figure \ref{Percolation} (b) shows that the particle did not percolate since there is no continuous connection of open sites from the top of the lattice, down to its bottom. Note that, in the illustration, von Neumann neighborhood was used.

We denote the percentage of open site vacancy of a system by $p$. The \textbf{percolation threshold} $p_{c}$ is the critical probability of open site vacancy for a system to percolate. That is, when the system has greater percentage of open-site occupation than the percolation threshold, there is an abrupt increase in the percolation probability. If $p>p_{c}$, the phase transition occurs and the system percolates. Otherwise, the system has low probability to successfully percolate.

\begin{figure}[H]
    \centering
    \includegraphics[scale=0.7]{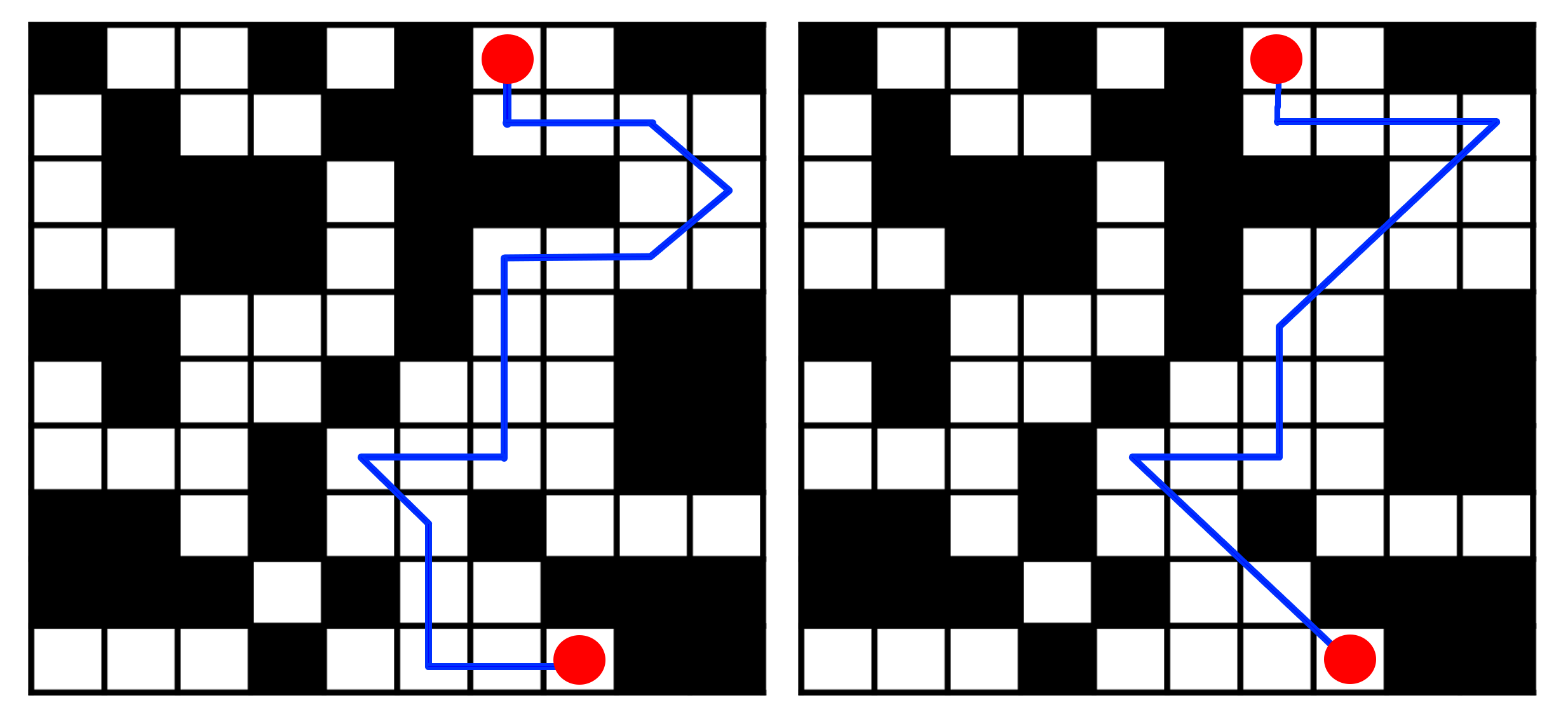}
    \caption{A configuration of the grid showing two different percolating paths (blue line) with the same initial and final positions}
    \label{fig:perc_path}
\end{figure}

Now, we define adjacency, path, and percolating path. Cells $x_{i,j}$ and $y_{k,l}$ are said to be \textbf{adjacent} if they are horizontally, vertically, or diagonally located beside each other (see Figure \ref{Neighborhood}). Furthermore, given cells $A(a_{i,j})$ and $B(b_{k,l})$, a \textbf{path} from $A$ to $B$ is a sequence of adjacent cells from $A$ to $B$, such that no cells are repeated. Lastly, if $A(a_{1,j})$ and $B(b_{N,l})$, $1\leq l \leq N$ are cells of an $N\times N$ grid, a path from $A$ to $B$ is a \textbf{percolating path}. Figure \ref{fig:perc_path} shows two different percolating paths for the same grid configuration.

There are cases that the barriers on a certain disordered grid may be governed by the concepts of tunneling and backscattering.

\textbf{Tunneling} is a quantum mechanical phenomenon in which a particle can pass through a potential barrier even classically, it does not have the energy to do so \cite{tunneling}. Particles can also \textbf{backscatter}. This is the reflection of a wave or the bounce back of a particle to its previous position when it hits a barrier.

The barrier can be an insulator, a vacuum, or a region of high potential energy. Moreover, Louis de Broglie proposed that the wavelength of a matter wave is inversely proportional to its velocity.

\begin{figure}[H]
    \centering
    \includegraphics[width = 0.55\textwidth]{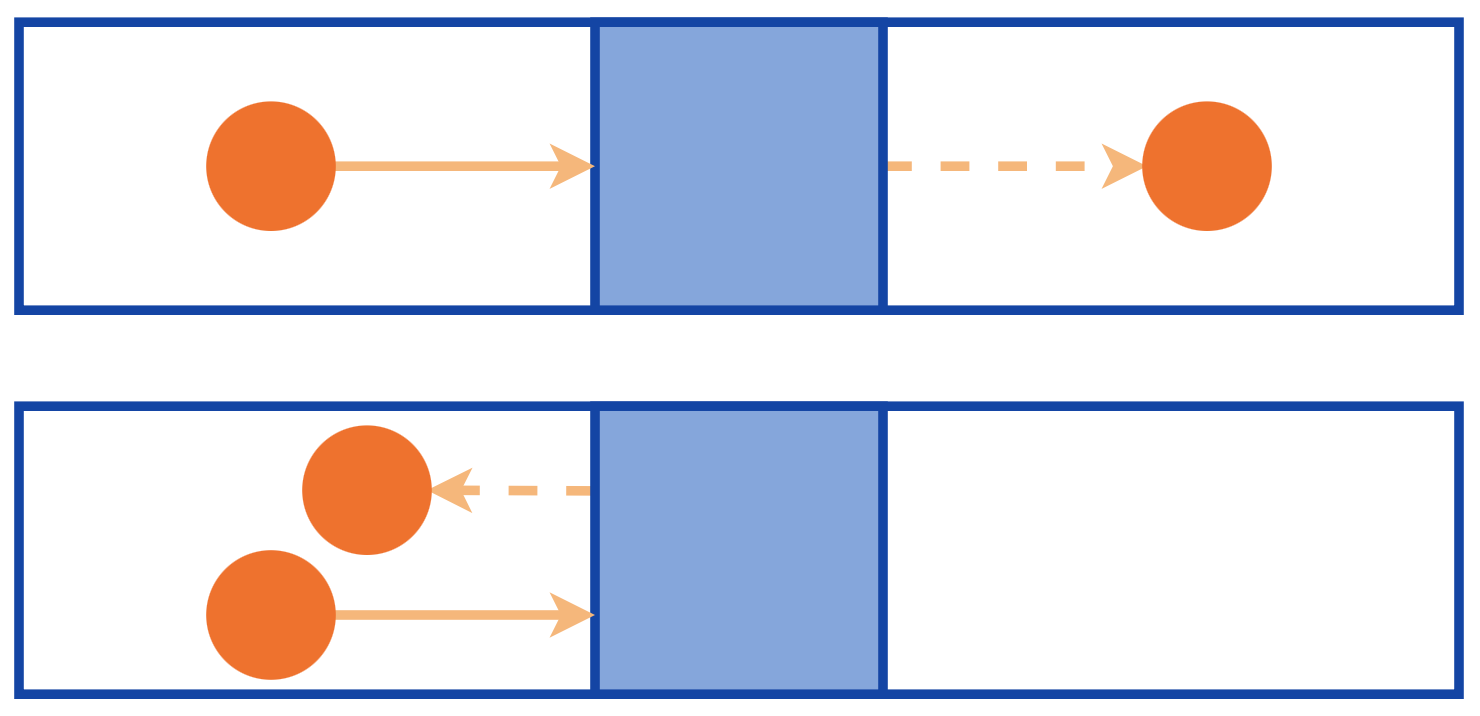}
    \caption{Tunneling (top) and backscaterring (bottom) phenomena for a particle. The dashed line indicates where tunneling or backscattering occurs.}
    \label{Tunneling}
\end{figure}

Figure \ref{Tunneling} (a) shows an illustration of a particle (orange) tunneling through a barrier (blue) that it cannot pass through classically and Figure \ref{Tunneling} (b) illustrates a particle that can backscatter. 

\section{The Cellular Automaton Model}
The cellular automaton (CA) model for particle transport is defined using its components as follows:

\begin{enumerate}
    \item Grid: $ D = N \times N $, where $ N \in \mathbb{N} $;
    \item Elementary States Set: $ E = \{ 0, 1, 2 \} $, where $ 0 $, $ 1 $, and $ 2 $ denote open sites, blocked sites, and filled sites (location of the particle), respectively;
    \item Neighborhood: Moore neighborhood, as shown in Figure \ref{Neighborhood};
    \item Local Rule $f$: depends on the conduction phenomena.
\end{enumerate}

\noindent Initial Condition: Randomize $x_{i,j}$ for $0 < i,j \leq N$.\\

Let $z^{t}(x_{i,j})$ be the state of the cell $x_{i,j}$ in iteration $t$; $p_{d}$ and $p_{di}$ are the probabilities for the particle to move downward and diagonally, respectively, with $p_{d}>p_{di}$. Moreover, let $p$, $q$, and $r$ be random numbers, and $p_{t}$ and $p_{b}$ are the tunneling and backscattering probabilities, respectively.\\

\noindent (a) Hopping Mechanism:

\begin{equation*}
    f(z^t|_{U(x_{i,j})})=
    \begin{cases}
        0, &\text{if } (p<p_{di} \text{ and } z^{t-1}(x_{i,j})=0) \text{ or } (z^{t-1}(x_{i,j})=0 \\
        & \quad \text{and } \forall x\in  U(x
        _{i,j}), z^{t-1}(x)\neq2), \\
        1, &\text{if } z^0(x_{i,j})=1, \\
        2, &\text{if } z^{t-1}(x_{i,j}) = 0 \text{ and } ((p\geq 1-p_d \text{ and } z^{t-1}(x_{i-1,j})=2), \text{ else if }\\ &\quad(p_{di}\leq p<1-p_d, \text{ and } z^{t-1}(x_{i-1,j\pm1})=2) \text{ else if } (z^{t-1}(x_{i,j\pm1})=2))
    \end{cases}
\end{equation*}

\noindent (b) Hopping Mechanism with Tunneling:
\begin{equation*}
    f(z^t|_{U(x_{i,j})})=
    \begin{cases}
        0, &\text{if } (p<p_{di} \text{ and } z^{t-1}(x_{i,j})=0) \text{ or } (z^{t-1}(x_{i,j})=0\\
        & \quad \text{and } \forall x\in  U(x
        _{i,j}), z^{t-1}(x)\neq2) \text{ or } p < 1- p_t, \\
        1, &\text{if } z^0(x_{i,j})=1, \\
        2, &\text{if } z^{t-1}(x_{i,j}) = 0 \text{ and } ((p\geq 1-p_d \text{ and } z^{t-1}(x_{i-1,j})=2), \text{ else if }\\ &\quad(p_{di}\leq p<1-p_d \text{ and } z^{t-1}(x_{i-1,j\pm1})=2), \text{ else if } (z^{t-1}(x_{i,j\pm1})=2)),\\
        & \quad\text{else if } q > 1-p_t \text{ and } i < N-1 \text{ and } ((p\geq 1 - p_d \text{ and }\\
        &\quad  z^{t-1}(x_{i-1,j}) = 1 \text{ and }z^{t-1}(x_{i-2,j}) = 2) \text{ or } (p_{di}\leq p\leq 1-p_d \text{ and }\\
        &\quad z^{t-1}(x_{i-1, j\pm 1} = 1) \text{ and }z^{t-1}(x_{i-2, j\pm 2}) = 2) \text{ or } (z^{t-1}(x_{i,j\pm1}) = 1\\
        & \quad \text{ and } z^{t-1}(x_{i,j\pm2}) = 2))
    \end{cases}
\end{equation*}
\noindent (c) Hopping Mechanism with Backscattering:
\begin{equation*}
    f(z^t|_{U(x_{i,j})})=
    \begin{cases}
        0, &\text{if } (p<p_{di} \text{ and } z^{t-1}(x_{i,j})=0) \text{ or } (z^{t-1}(x_{i,j})=0\\
        & \quad \text{and } \forall x\in  U(x
        _{i,j}), z^{t-1}(x)\neq2) \text{ or } p < 1- p_b, \\
        1, &\text{if } z^0(x_{i,j})=1, \\
        2, &\text{if } z^{t-1}(x_{i,j}) = 0 \text{ and } ((p\geq 1-p_d \text{ and } z^{t-1}(x_{i-1,j})=2), \text{ else if }\\ &\quad(p_{di}\leq p<1-p_d \text{ and } z^{t-1}(x_{i-1,j\pm1})=2), \text{ else if } (z^{t-1}(x_{i,j\pm1})=2)),\\
        & \quad\text{else if } r > 1-p_b \text{ and } i>1 \text{ and } ((p\geq 1 - p_d \text{ and}\\
        &\quad z^{t-1}(x_{i+2,j}) = 1 \text{ and } z^{t-1}(x_{i+1,j} = 2)) \text{ or } (p_{di}\leq p\leq 1-p_d \text{ and } \\
        &\quad z^{t-1}(x_{i+2, j\pm 2} = 1) \text{ and } z^{t-1}(x_{i+1, j\pm 1}) = 2) \text{ or } (z^{t-1}(x_{i,j\pm1}) = 1 \text{ and }\\
        & \quad z^{t-1}(x_{i,j\pm2}) = 2))
    \end{cases}
\end{equation*}

\noindent From the CA model above, we can deduce the following properties.\\

\noindent \textbf{Property 1:} For a fixed $p$, if $z^{0}(x_{i,j})=1,$ then $z^{t}(x_{i,j})=1,$ for all $t.$\\

In the following, we consider the extreme case that all cells of the grid are blocked:

\noindent \textbf{Property 2:} If $z^{0}(x_{i,j})=1$, for all $ 1 \leq i,j \leq N, $ then $z^{t}(x_{i,j})=1$, for all $ 1 \leq i,j \leq N $ and $t.$\\

\noindent\textbf{Note:} The properties above can easily be deduced from the definition of the model.\\

\noindent \textbf{Theorem:} If $z^{0}(x_{i,j})=0$, for all $ 1 \leq j \leq N, $ then $z^{t}(x_{i,j})=2,$ for some $j$ and $t$.
\begin{proof}
We shall recall $p$ and $p_{c}$ to be the percentage of open site vacancy, and the percolation threshold of the system. Assume that $z^{0}(x_{i,j})=0$, for all $ 1 \leq j \leq N. $ Then, $p=1>p_{c}.$ Since $p>p_{c},$ the system percolates, i.e., $$z^{t}(x_{i,j})=2,$$ for some $j$ and $t$.
\end{proof}

\begin{remark}
In the preceding theorem, if all cells are open, we are sure that the system percolates.
\end{remark}

The success of a particle to percolate can be summarized in an algorithm as follows:\\

Suppose that for a fixed $p$, all probability conditions of the CA model are met. Assume that the particle is located at the cell $x_{1,j}$, for some $1\leq j \leq N$ at iteration $t=0$. Then, for $i \geq 1$,\\

\noindent\textit{For Hopping Mechanism:}
\begin{enumerate}
    \item If $z^{t}(x_{i+1,j})=0$, then $z^{t+1}(x_{i+1,j})=2$.
    \item Else if $z^{t}(x_{i+1,j\pm1})=0$, then $z^{t+1}(x_{i+1,j\pm1})=2$.
    \item Else if $z^{t}(x_{i,j\pm1})=0$, then $z^{t+1}(x_{i,j\pm1})=2$.
    \item Repeat the process until the particle is located at $x_{N,j}$, for some $j$ and $t$.
\end{enumerate}
\textit{For Hopping Mechanism with Tunneling:}
\begin{enumerate}
    \item If $z^{t}(x_{i+1,j}) = 0$, then $z^{t+1}(x_{i+1,j})=2$.
    \item Else if $z^{t}(x_{i+1,j\pm1})=0$, then $z^{t+1}(x_{i+1,j\pm1})=2$.
    \item Else if $z^{t}(x_{i,j\pm1})=0$, then $z^{t+1}(x_{i,j\pm1})=2$.
    \item Else if $i<N-1$, $z^{t}(x_{i+1,j})=1$, and $z^{t}(x_{i+2,j})=0$, then $z^{t+1}(x_{i+2,j})=2$.
    \item Else if $i<N-1$, $z^{t}(x_{i+1,j\pm1})=1$, and $z^{t}(x_{i+2,j\pm2})=0$, then $z^{t+1}(x_{i+2,j\pm2})=2$.
    \item Else if $z^{t}(x_{i,j\pm1})=1$, and $z^{t}(x_{i,j\pm2})=0$, then $z^{t+1}(x_{i,j\pm2})=2$.
    \item Repeat the process until the particle is located at $x_{N,j}$, for some $j$ and $t$.
\end{enumerate}
\textit{For Hopping Mechanism with Backscattering:}
\begin{enumerate}
    \item If $z^{t}(x_{i+1,j})=0$, then $z^{t+1}(x_{i+1,j})=2$.
    \item Else if $z^{t}(x_{i+1,j\pm1})=0$, then $z^{t+1}(x_{i+1,j\pm1})=2$.
    \item Else if $z^{t}(x_{i,j\pm1})=0$, then $z^{t+1}(x_{i,j\pm1})=2$.
    \item Else if $i>1$, $z^{t}(x_{i+1,j})=1$, and $z^{t}(x_{i-1,j})=0$, then $z^{t+1}(x_{i-1,j})=2$.
    \item Else if $i>1$, $z^{t}(x_{i+1,j\pm1})=1$, and $z^{t}(x_{i-1,j\mp1})=0$, then $z^{t+1}(x_{i-1,j\mp1})=2$.
    \item Else if $z^{t}(x_{i,j\pm1})=1$, and $z^{t}(x_{i,j\mp1})=0$, then $z^{t+1}(x_{i,j\mp1})=2$.
    \item Repeat the process until the particle is located at $x_{N,j}$, for some $j$ and $t$.
\end{enumerate}

\section{Simulation and results}
\hspace{\parindent}
In this section, the cellular automaton model is implemented using computer simulation. Then, we investigate the effects of lattice sizes and different conduction phenomena in the behavior of percolation a posteriori probability across different open-site vacancies. We also present some mathematical properties of the model by considering some trivial cases taken from the computer simulation, and constructing some ordered grid configurations and examine their different properties such as number of percolating paths and percolation success rates.

\subsection{Computer Simulation for Disordered Systems}
\hspace{\parindent}

The percolation probability versus percentage of open-site vacancy graphs for hopping mechanism, hopping mechanism with tunneling and/or backscattering emulate an exponential behavior. For this simulation, we let $ N=10, \quad p_{d}=50\%,\quad p_{di}=25\%, \quad p_{t}=50\%, \quad \text{and} \quad p_{b}=50\% $. For the graphs, the $x$- and $y$- axes denote the percentage of open site vacancy and the percolation probability, respectively.

\begin{figure}[H]
    \centering
    \includegraphics[scale=0.6]{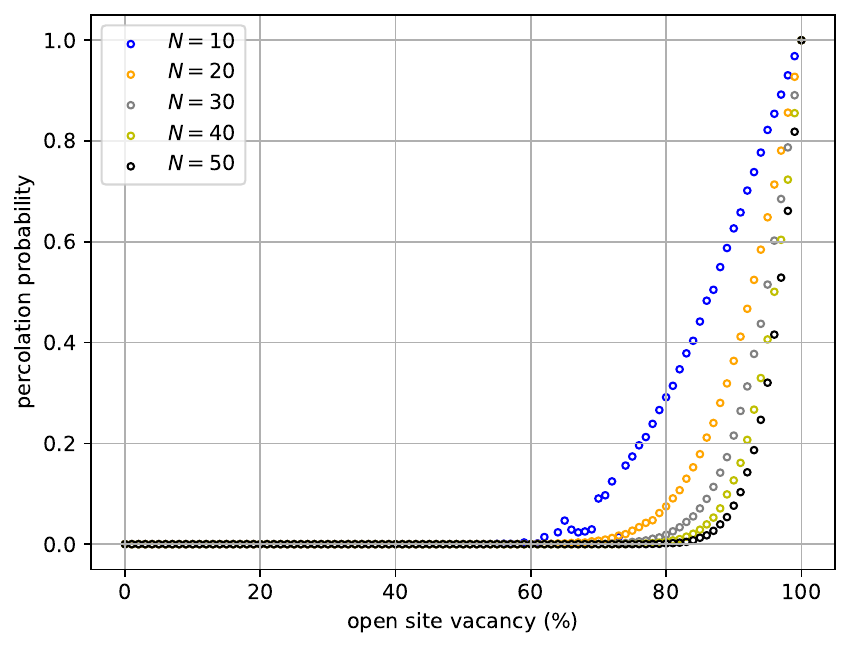}
    \caption{Combined Percolation Probability versus Open-site Vacancy graphs for Hopping Mechanism ($N=10, 20, 30, 40, 50$)}
    \label{CHMGraph}
\end{figure}

Figures \ref{CHMGraph}--\ref{CHMBGraph} below are the combinations of percolation probability versus percentage of open site vacancy graphs for hopping mechanism, hopping mechanism with tunneling, hopping mechanism with backscattering, and hopping mechanism with tunneling and backscattering for grid sizes $N=10, 20, 30, 40, 50$, respectively.

\begin{figure}[H]
    \centering
    \includegraphics[scale=0.6]{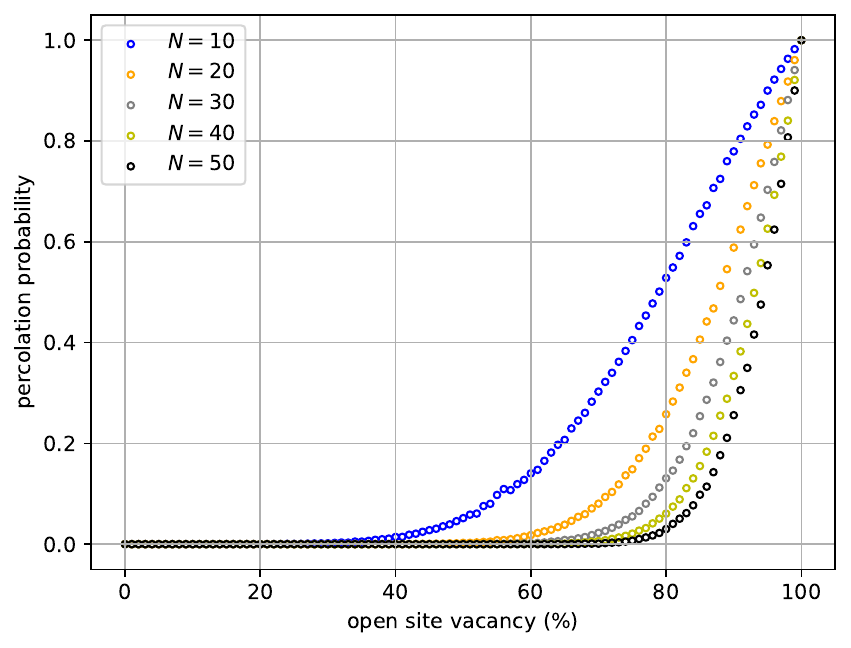}
    \caption{Combined Percolation Probability versus Open-site Vacancy graphs for Hopping Mechanism with Tunneling ($N=10, 20, 30, 40, 50$)}
    \label{CHMTGraph}
\end{figure}

\begin{figure}[H]
    \centering
    \includegraphics[scale=0.6]{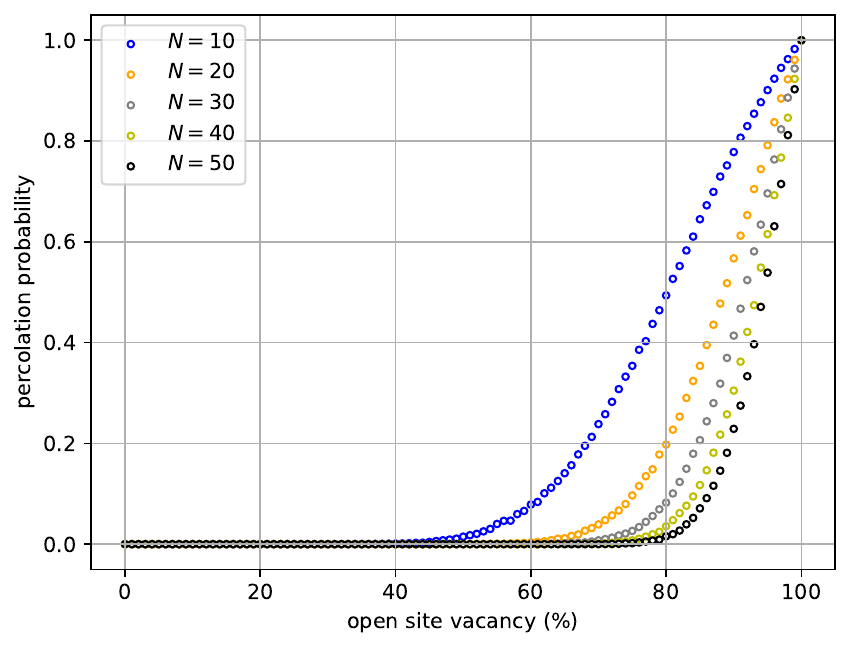}
    \caption{Combined Percolation Probability versus Open-site Vacancy graphs for Hopping Mechanism with Backscattering ($N=10, 20, 30, 40, 50$)}
    \label{CHMBGraph}
\end{figure}

\begin{figure}[H]
    \centering
    \includegraphics[scale=0.6]{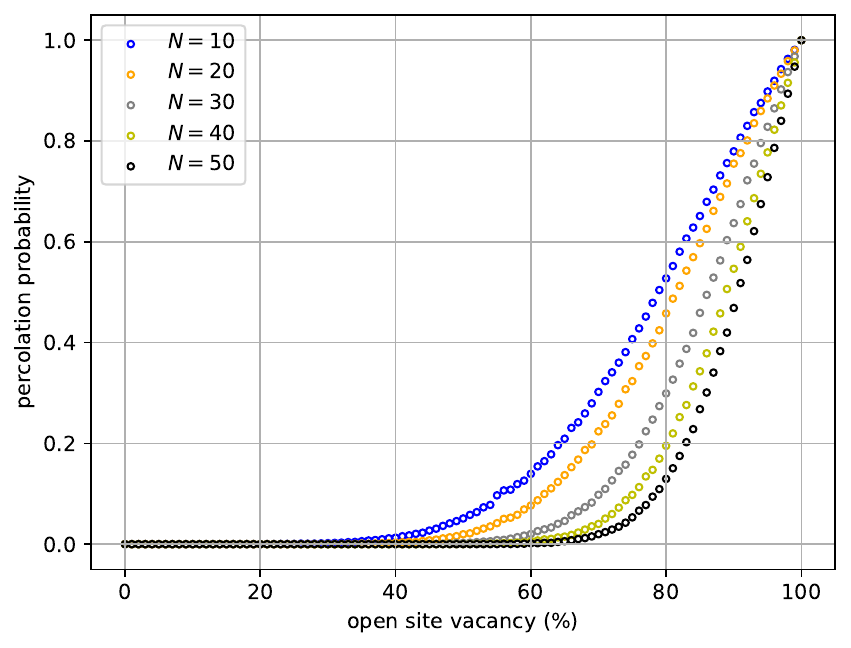}
    \caption{Combined Percolation Probability versus Open-site Vacancy graphs for Hopping Mechanism with Tunneling and Backscattering ($N=10, 20, 30, 40, 50$)}
    \label{CHMBGraph}
\end{figure}

It can be observed that as the grid size increases, the onset to percolate of the graph is getting displaced to a larger value. However, the property of the system to percolate at $p=100\%$ is not affected by the change in grid size. The onset is also displaced into a smaller open-site vacancy for the cases of hopping mechanism with tunneling or backscattering compared to hopping mechanism alone. From these graphs, it can be inferred that the model is scale-dependent.

Table 1 in Appendix \ref{sec:appendix-ordered} shows that the percolation success rates of the ordered systems are generally greater than their disordered counterpart, based on their grid size.

\begin{figure}[H]
    \centering
    \includegraphics[scale=0.6]{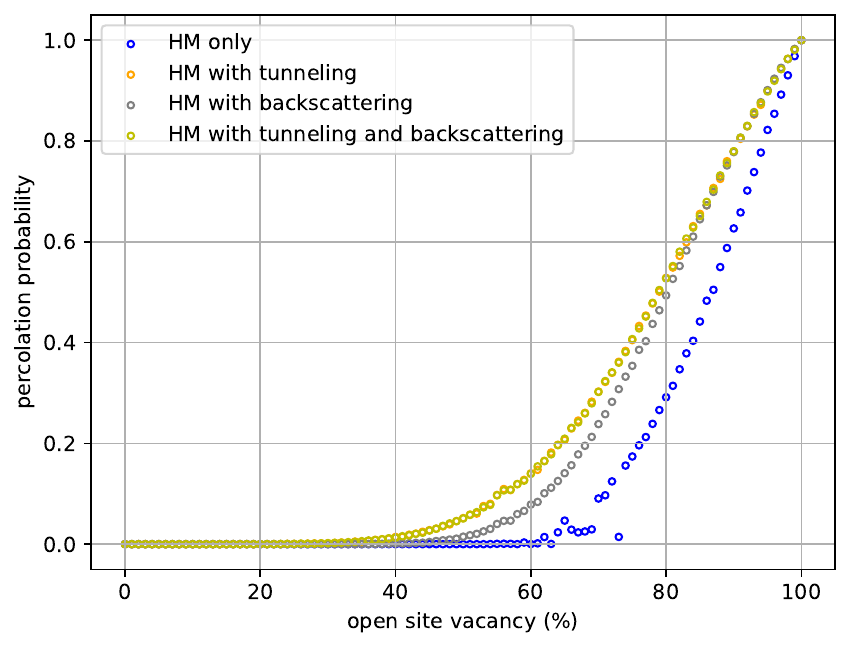}
    \caption{Combined Percolation Probability versus Open-site Vacancy graphs for Hopping Mechanism (HM) alone, Hopping Mechanism with Tunneling, Hopping Mechanism with Backscattering, and Hopping Mechanism with Tunneling and Backscattering ($N=10$)}
    \label{CHMBGraph2}
\end{figure}

From Figure \ref{CHMBGraph2}, we can observe that the hopping mechanism with tunneling and hopping mechanism with tunneling and backscattering graphs are somewhat identical. These cases have the greatest effect from the case of hopping mechanism alone compared to hopping mechanism with backscattering.

According to Quantum theory, the behavior of the percolation probability of a particle in a system consisting of varying number of potential barriers, with a certain tunneling probability, imitates an exponential behavior of the form $$ \phi (p)=Ae^{kp}, $$ where $A,k$ are real constants and $p$ is the open-site vacancy of the system. When the data from Figure 5.5 were subjected to exponential fitting, the resulting equation is $y=0.011233e^{0.0462847p}$, with $r^{2}=0.9714$, $r$ being the Pearson's correlation coefficient. On the other hand, its sigmoidal fitting gives $y=\frac{0.414618}{1+e^{-0.0208228x}}$ with $r^{2}=0.1736$. From this, we can infer that the data fits an exponential behavior, but somehow still have a sigmoidal behvior. The computational results in this study follow the behavior from phenomenological physical models and experiments \cite{Balberg}.

\section{Conclusion and Open Problems}
\hspace{\parindent}
This work constructed a cellular automaton model specific for particle transport with different conduction phenomena: tunneling and backscattering. The three models have separate local rules but are identical in the type of grid, neighborhood and elementary state used. We also presented three algorithms for finding the percolating path on any 2-dimensional disordered system, specific for each case: hopping mechanism, hopping mechanism with tunneling, and hopping mechanism with backscattering. 

A computer simulation was also developed for the constructed cellular automaton model. From the simulations of $30,000$ runs, with grid size $N=10,20,30,40,50$. The computer simulation was verified to work properly using ordered systems that we constructed. We also investigated the open-site vacancy, number of percolating paths, and percolation success rate based on the CA model computer simulation for hopping mechanism and hopping mechanism with tunneling. 

It was observed that the model is scale-dependent, that is, as we increase the grid size, the onset of the percolation probability versus open-site vacancy is displaced on a larger value. Moreover, the behavior of the percolation probability versus open-site vacancy for hopping mechanism with tunneling fits an exponential function, which agrees with the theory associated for systems with tunneling.

Theoretically, it was proven that for $p=100\%$, or if there are no barriers in the system, the point particle will successfully percolate $100\%$ of the time. On the other hand, if $p=0\%$, or there are no open sites in the system, any particle cannot pass through the system. We also observed that hopping mechanism with tunneling and hopping mechanism with tunneling and backscattering have the greatest effect on the disordered system.

The percolation success rates of these grids are greater than the percolation success rates of their random and disordered configurations counterpart, based on grid size.

For future works, one may try to develop a computer simulation for the model that will compute for the displacement corresponding to each open-site vacancy and investigate the effect of grid size on the displacement of the particle on the grid. This can help for analysis on which point in the grid the particle stops. One may also include computation of the time parameter and see how long, on average, it takes the particle to hop onto the bottom from the top of the grid. Another case to consider is extending the dependence of the tunneling probability on the width of the barrier faced by the particle along its way to the bottom of the grid.

Another topic for the expansion of this study is the construction of another cellular automaton model which incorporates the temperature dependence of the current model presented in this study. Moreover, future studies may also include presentation of more ordered grid configurations, investigate their properties for physical applications of the model, and characterize the CA model.

\begin{appendices}
\section{Computer Simulation for Ordered Systems} \label{sec:appendix-ordered}

\hspace{\parindent}
In this section, we verify that the computer simulation works properly using some ordered systems. We constructed different ordered configurations of the grid and computed the number of percolating paths for each configuration, for any $N$. Other factors like success rates in the computer simulations were also investigated. \\

\begin{figure}[H]
    \hspace{-1.5cm}
    \includegraphics[scale=0.8]{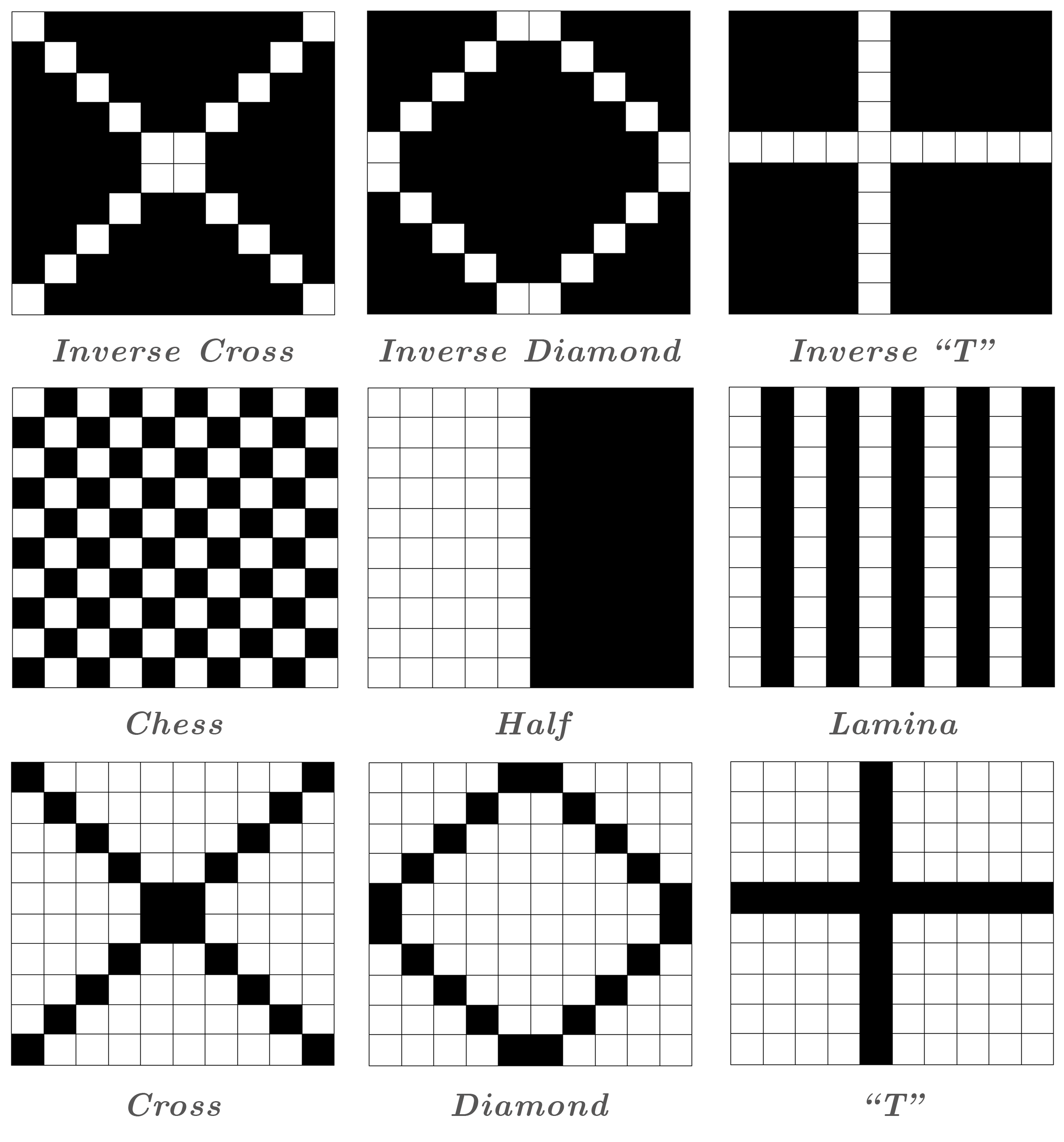}
    \label{config}
\end{figure}
\begin{figure}[H]
    \hspace{-3cm}
    \includegraphics[scale=1.1]{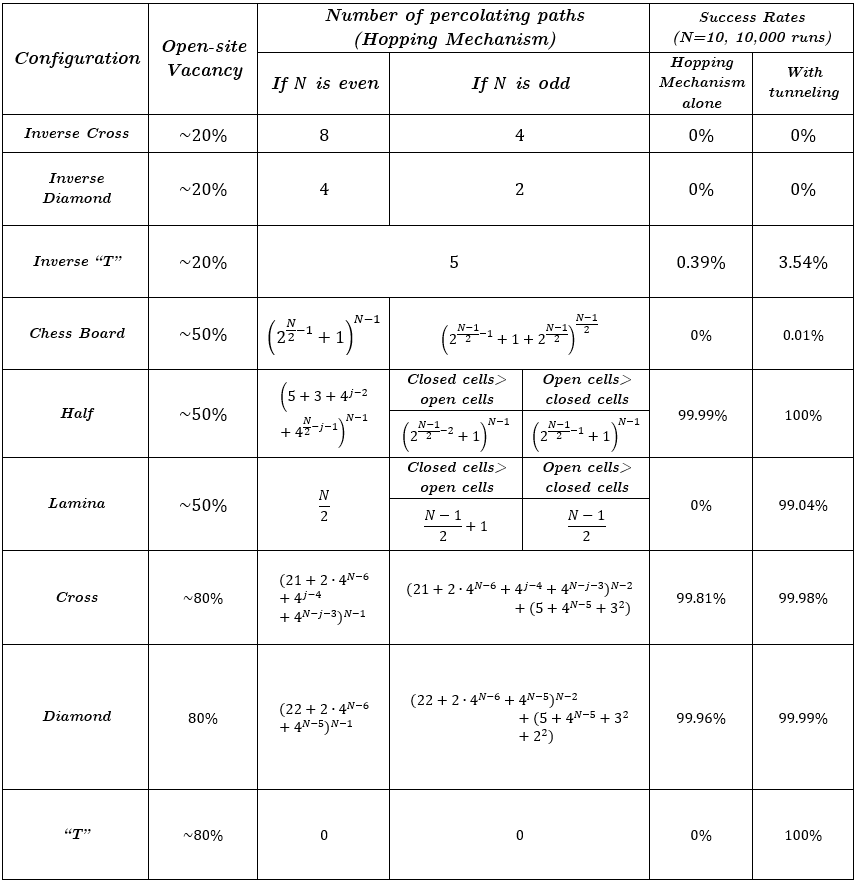}
    \label{table}
\end{figure}
\begin{center}
    \vspace{-0.5cm}
    Table 1. Open-site vacancy and number of percolating paths for some ordered configuration of an $N \times N$ grid
    \vspace{1cm}
\end{center}

In Table 1, we can observe that as we increase the percentage of open site vacancy of the system, the number of percolating paths also increases, and the generally, the percolation success rates follow this trend.

To get the number of percolating paths, consider an $N \times N$ grid with \textit{chess} configuration, $N$ is even. We count the number of percolating paths of the said configuration of grid by exhausting all the possible paths.

In any row, there are $\frac{N}{2}$ open cells. Note that one of those open cells is placed on either left- or rightmost cell of the row. 

Hence, $\frac{N}{2}-1$ cells will have two choices as possible destination (i.e., diagonal left or diagonal right), while the other one has only once choice.

This scenario is repeated for $N-1$ times since when the particle reaches the bottom row, the algorithm stops. 

So, the total number of percolation paths is given by $$\Big[ 2^{\frac{N}{2}-1} +1 \Big]^{N-1}$$.

If $N$ is odd, cases in a row are further divided into two: number of closed cells are greater than that of the open cells and vise versa.

This method is followed to get the number of percolating paths for all other systems of ordered configurations.

\end{appendices}

\bibliographystyle{plain}
\bibliography{biblio}

\end{document}